\documentclass{jfm}

\usepackage{graphicx}
\usepackage{natbib}

\ifCUPmtlplainloaded \else
  \checkfont{eurm10}
  \iffontfound
    \IfFileExists{upmath.sty}
      {\typeout{^^JFound AMS Euler Roman fonts on the system,
                   using the 'upmath' package.^^J}%
       \usepackage{upmath}}
      {\typeout{^^JFound AMS Euler Roman fonts on the system, but you
                   dont seem to have the}%
       \typeout{'upmath' package installed. JFM.cls can take advantage
                 of these fonts,^^Jif you use 'upmath' package.^^J}%
      }
  \else
  \fi
\fi

\ifCUPmtlplainloaded \else
  \checkfont{msam10}
  \iffontfound
    \IfFileExists{amssymb.sty}
      {\typeout{^^JFound AMS Symbol fonts on the system, using the
                'amssymb' package.^^J}%
       \usepackage{amssymb}%

      }{}
  \fi
\fi

\ifCUPmtlplainloaded \else
  \IfFileExists{amsbsy.sty}
    {\typeout{^^JFound the 'amsbsy' package on the system, using it.^^J}%
     \usepackage{amsbsy}}
    {}
\fi

\newsavebox{\astrutbox}
\sbox{\astrutbox}{\rule[-5pt]{0pt}{20pt}}

\title[Numerical simulation of super-square patterns in Faraday waves]{Numerical simulation of super-square patterns in Faraday waves}

\author[L.~Kahouadji, N.~P\'erinet, L.S.~Tuckerman, S.~Shin, J.~Chergui and D.~Juric]%
{L.~Kahouadji$^{1,2}$,\ns  N.~P\'erinet$^3$, L.S.~Tuckerman$^1$
\thanks{Email address for correspondence: laurette@pmmh.espci.fr},\ns S.~Shin$^4$, J.~Chergui$^5$ and D.~Juric$^5$}%

\affiliation{
$^1$PMMH (UMR 7636 CNRS - ESPCI - UPMC Paris 6 - UPD Paris 7 - PSL),
10 rue Vauquelin, 75005 Paris France\\[\affilskip]
$^2$Department of Chemical Engineering, Imperial College London, South Kensington Campus, London SW7 2AZ, United Kingdom\\[\affilskip]
$^3$Departamento de F\'isica, Facultad de Ciencias F\'isicas y Matem\'aticas, Universidad de Chile, Santiago, Chile\\[\affilskip]
$^4$Department of Mechanical and System Design Engineering, Hongik University, Seoul 121-791, Republic of Korea\\[\affilskip]
$^5$LIMSI-CNRS, B\^at 508, rue John von Neumann - 91405 Orsay, France}

\pubyear{2015}
\volume{650}
\pagerange{119--126}
\date{?; revised ?; accepted ?.}
\begin{document}

\maketitle

\begin{abstract}
We report the first simulations of the Faraday instability using the
full three-dimensional Navier-Stokes equations in domains much larger
than the characteristic wavelength of the pattern. We use a massively
parallel code based on a hybrid Front-Tracking/Level-set algorithm for
Lagrangian tracking of arbitrarily deformable phase
interfaces. Simulations performed in square and cylindrical domains
yield complex patterns. In particular, a superlattice-like pattern
similar to those of [Douady \& Fauve, {\em Europhys. Lett.}~{\bf 6},
  221-226 (1988); Douady, {\em J. Fluid Mech.}~{\bf 221}, 383-409
  (1990)] is observed. The pattern consists of the superposition of
two square superlattices.  We conjecture that such patterns are
widespread if the square container is large compared to the critical
wavelength.  In the cylinder, pentagonal cells near the outer wall
allow a square-wave pattern to be accommodated in the center.

\end{abstract}

%\begin{keywords}
%05 65.+b, Self-organized systems, 05 45.-a, Nonlinear dynamics and chaos, 47.20.Ma Interfacial instabilities , 47.35.Pq, Capillary waves
%\end{keywords}
%\begin{keywords}
%Faraday waves, pattern formation
%\end{keywords}

\section{\label{sec:level1}Introduction}

In the \cite{Faraday-ptrsl-1831} problem, the interface between two
superposed fluid layers, subjected to periodic vertical vibration of
sufficient amplitude, forms sustained standing wave patterns, called
Faraday waves.  Many universal dynamical-systems phenomena were first
discovered through the Faraday experiment.  In particular, a number of
experimental investigations have focused on the variety of patterns
formed in square
\cite[e.g.][]{Douady-epl-1988,Douady-jfm-1990,Simonelli-jfm-1989} and
cylindrical \cite[e.g.][]{Ciliberto-jfm-1985,Das-jfm-2008,Batson-jfm-2013,Rajchenbach-prl-2013}
containers, which have led to important theoretical developments
\cite[e.g.][]{Meron-pra-1987,Silber-phl-1989,Crawford-physd-1990,Crawford-physd-1991,Gomes-nonl-1994}
in the fields of nonlinear physics and pattern formation.

Recently, \cite{Perinet-jfm-2009,Perinet-prl-2012}
performed the first three-dimensional direct numerical simulation of
Faraday waves in horizontally periodic domains containing a few basic
cells.  Here, we describe numerical simulations of Faraday waves in
much larger domains which have yielded more complicated patterns. For 
this purpose, we have used the free-surface code developed by
\cite{Shin-cf-2014} for simulations of two-phase flows on massively
parallel computer architectures.  
In a square domain (i.e.~with square horizontal cross-section), we
have obtained patterns with two different length scales, 
very similar to those reported in
experiments by \cite{Douady-epl-1988} and \cite{Douady-jfm-1990}. 
Patterns of this type were later termed superlattices by 
\cite{Kudrolli-physd-1998} and \cite{Arbell-pre-2002}, 
who observed a large variety of such patterns in their experiments, 
but by using a temporal forcing with two frequencies.
The spectral analysis of our patterns, produced by
single-frequency forcing, shows that they result primarily from the
superposition of two superlattices constructed from very similar
wavelengths.
We have also simulated the same conditions in a cylindrical
container. The resulting interface contains squares and pentagons.

The paper is organized as follows. First, we briefly present the
problem and explain the methods involved in the code.  We then present
the numerical linear stability analysis that is validated by
comparison with the theoretical treatment of \cite{Kumar-jfm-1994}. We
demonstrate the robustness of the obtained superlattice-like pattern
by testing several boundary conditions. This
pattern is described and analysed spectrally. Then, we show the
results of simulations with the same conditions but in a cylindrical
domain.

\section{Problem formulation, governing equations and numerical scheme}

The governing equations for an incompressible two-phase flow can be
expressed by a single field formulation:
\begin{equation}\label{NS-Eqs}
\displaystyle{\rho  \left( \frac{\partial \textbf{u}}{\partial t}
+\textbf{u}\cdot\nabla \textbf{u}\right) =  -\nabla P + \rho \textbf{G}
+ \nabla \cdot \mu\left(\nabla\textbf{u} +\nabla\textbf{u}^T \right)
+ \textbf{F}}, \qquad \displaystyle{\nabla\cdot\textbf{u}=0 }
\end{equation}
where $\textbf{u}$ is the velocity, $P$ is the pressure, $\rho$ is the
density, $\mu$ is the dynamic viscosity, $\textbf{G}$ represents 
the homogeneous volume forces, and $\textbf{F}$ is the local surface
tension force at the interface. Here, ${\bf G}$ contains the
gravitational term supplemented by a time-dependent force
accounting for the vibrations of the domain:
\begin{equation}\label{forcing}
{\bf G}=-\left(g+\gamma\cos(\omega t)\right)\textbf{e}_z
\label{grav}\end{equation}
where $g$ is the gravitational acceleration, 
$\textbf{e}_z$ is the upward vertical unit vector, 
$\gamma$ is the amplitude of the inertial forcing 
and $\omega=2\pi/T$ its frequency.

The code essentially consists of two modules: one, which solves the
three-dimensional incompressible Navier-Stokes equations, and the
other, which treats the free surface using a parallel Lagrangian
Tracking method and is only active if the flow is a two-phase flow.
Material properties such as density or viscosity are defined in the
entire domain:
\begin{equation}
\left.\begin{array}{c}
\rho(\textbf{x},t) = \rho_{1} +\left(  \rho_{2} -\rho_{1}\right)I(\textbf{x},t)\\
\mu(\textbf{x},t) = \mu_{1} +\left( \mu_{2} -\mu_{1}\right)I(\textbf{x},t).
\end{array}\right.
\end{equation}
The indicator function, $I$, is a numerical Heaviside
function, ideally zero in one phase and one in the other phase. $I$ is
resolved with a sharp but smooth transition across 3 to 4 grid cells
and is generated using a vector distance function computed directly
from the tracked interface (\cite{Shin-ijnmf-2009}).

The fluid variables $\textbf{u}$ and $P$ are calculated by a
projection method 
%%(\cite{Chorin-mc-1968,Goda-jcp-1979}). 
\cite[][]{Chorin-mc-1968}. 
The temporal scheme is first order, 
with implicit time integration used for the viscous terms.
For spatial discretization we use the staggered-mesh
marker-in-cell (MAC) method
\cite[][]{Harlow-pof-1965} on a uniform finite-difference grid 
with second-order essentially non-oscillatory (ENO) advection
%(\cite{Shu-jcp-1989,Sussman-cf-1998}). 
\cite[][]{Shu-jcp-1989}.
The pressure and
distance function are located at cell centers while the $x$, $y$ and
$z$ components of velocity are located at the faces. All spatial
derivatives are approximated by standard second-order centered
differences. The treatment of the free surface uses a hybrid
Front-Tracking/Level-Set technique which defines the interface both by
a discontinuous density field on the Eulerian grid as well as by 
triangles on the Lagrangian interface mesh.

The surface tension $\textbf{F}$ is implemented by the hybrid formulation
%(\cite{Shin-jcp-2005})
\begin{equation}
\textbf{F} = \sigma \kappa_{H} \nabla I , \qquad
\kappa_{H}=\frac{\textbf{F}_{L}\cdot\textbf{N}}{\textbf{N}\cdot\textbf{N}}\end{equation}
where $\sigma$ is the surface tension coefficient assumed to be
constant and $\kappa_{H}$ is twice the mean interface curvature
field calculated on the Eulerian grid, with
\begin{equation}
\textbf{F}_{_{L}} = \int_{\Gamma(t)} \kappa_f  \textbf{n}_{f} \delta_f
\left( \textbf{x} - \textbf{x}_{f} \right) \mathrm{d}s ,
\qquad \textbf{N} = \int_{\Gamma(t)}  \textbf{n}_{f} \delta_f
\left( \textbf{x} - \textbf{x}_{f} \right) \mathrm{d}s
\end{equation}
Here, $\textbf{x}_f$ is a parameterization of the time-dependent
interface, $\Gamma(t)$, and
$\delta_f \left( \textbf{x} -\textbf{x}_{f} \right) $
is a Dirac distribution that is non-zero only
where $\textbf{x}=\textbf{x}_f$; $\textbf{n}_f$ stands for the unit
normal vector to the interface and $ \mathrm{d}s$ is the length of an
interface element; $\kappa_f$ is twice the mean interface curvature
obtained on the Lagrangian interface. The geometric information, unit
normal, $\textbf{n}_f$, and interface element length, $ \mathrm{d}s$
in $\textbf{N}$ are computed directly from the Lagrangian interface
and then distributed onto the Eulerian grid using the discrete delta
function and the immersed boundary method of \cite{Peskin-jcp-1977}. A
detailed description of the procedure for calculating $\textbf{F}$,
$\textbf{N}$ and $I$ can be found in \cite{Shin-jcp-2007}.

The Lagrangian interface is advected by integrating
$\mathrm{d}\textbf{x}_f/\mathrm{d}t = \textbf{V}$ with a
second-order Runge-Kutta method where the interface velocity,
$\textbf{V}$, is interpolated from the Eulerian velocity.

The time step $\Delta t$ is chosen at each iteration in order to satisfy a criterion based on
\begin{equation}\label{steps}
\left\{\Delta t_{\rm CFL}, \Delta t_{\rm int}, \Delta t_{\rm vis}, \Delta t_{\rm cap}\right\}
\end{equation}
which ensures stability of the calculations.  These bounds are defined by:
\begin{equation}\label{Time_steps}
\begin{array}{ll}
\displaystyle{\Delta t_{\rm CFL} \equiv \min_j \left( \min_{\rm domain} \left(\frac{\Delta x_j}{u_j}\right) \right)} &
\displaystyle{\Delta t_{\rm int} \equiv \min_j \left( \min_{\Gamma(t)} \left( \frac{\Delta x_j}{\|{\bf V}\|}\right) \right)}\\
\displaystyle{\Delta t_{\rm vis} \equiv \min\left(\frac{\rho_{2}}{\mu_{2}},\frac{\rho_{1}}{\mu_{1}}\right)\frac{\Delta x_{\min}^2}{6}}&
\displaystyle{\Delta t_{\rm cap} \equiv \frac{1}{2}\bigg(\frac{(\rho_{1} + \rho_{2}) \Delta x_{\min}^3}{\pi \sigma} \bigg)^{1/2}}
\end{array}
\label{eq:bounds}\end{equation}
where $\Delta x_{\min} = \min_j \left(\Delta x_j \right)$.
With the periodic volume force term $\textbf{G}$ of (\ref{forcing}),
a supplementary upper bound $\Delta t_{\omega}=2\pi/(50\omega)$ is required.

Fortran 2003 allows the definition of a set of dynamically allocated
derived types and generic procedures associated with the 
grids, scalar and vector fields, operators as well as the various solvers 
used in the Navier-Stokes and Lagrangian tracking modules. 
The parallelization of the code is based on algebraic domain
decomposition, where the velocity field is solved by a parallel 
generalized minimum residual (GMRES)
method for the implicit viscous terms and the pressure by a parallel
multigrid method motivated by the algorithm of
\cite{Kwak-InterScience-2004}.  Communication across process threads
is handled by message passing interface (MPI) procedures.

Finally, the code also contains a module for the definition of
immersed solid objects and their interaction with the flow, 
which we have used to simulate Faraday waves in a cylindrical container.
A Navier-slip dynamic contact line model is implemented where 
hysteresis is accounted for by fixing the contact angle limits to
prescribed advancing or receding angles as in \cite{Shin-jmst-2009}.
Further details are available in \cite{Shin-cf-2014}.

\section{Parameters and linear stability}
\label{param}

\begin{figure*}
\includegraphics*[width=\columnwidth]{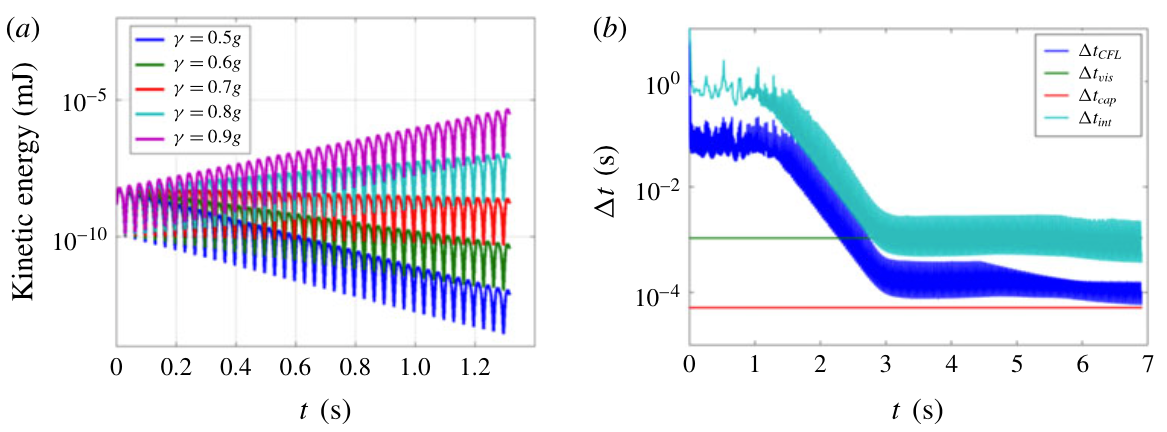}
\caption{\label{Floquet} a) Evolution of the total kinetic energy for $\omega/2\pi=30$ Hz and various accelerations in a small domain. The fact that the peaks for a given acceleration lie along a line shows that the energy growth is very close to exponential. b) Evolution of the various bounds on the time step
during the course of the simulation. The smallest, $\Delta t_{\rm cap}$, is used in the simulations. }
\end{figure*}

\begin{table}
\begin{center}
\begin{tabular}{ccccc}
$\omega/2\pi $& $\lambda_c$ & Floquet & DNS & relative error\\
Hz & mm & $\gamma_c/g$ & $\gamma_c/g$ & Floquet/DNS\\
\hline
30 & 11.3 & 0.733 & 0.713 & 2.73 \% \\
60 & 5.9 & 2.65 & 2.61 & 1.50 \%\\
90 & 4.3 & 5.32 & 5.3 & 0.38 \%\\
\end{tabular}
\end{center}
\caption{\label{tab_threshold}Comparison of Floquet and DNS instability thresholds for various frequencies.}
\end{table}

We consider a layer of silicone oil of thickness $h=14.5$ mm with
density $\rho_{1} = 965$ kg m$^{-3}$ and dynamic viscosity $\mu_{1} =
0.02$ kg m$^{-1}$s$^{-1}$. This layer is overlaid with $29.5$ mm of
air of density $\rho_{2}= 1.205$ kg m$^{-3}$ and dynamic viscosity
$\mu_{2}=1.82\times10^{-5}$ kg m$^{-1}$s$^{-1}$. The surface tension
at the interface between the oil and the air is $\sigma=0.02$ kg
s$^{-2}$.
%LST added
This choice of parameters was originally motivated by experiments on 
Faraday waves in a rotating cylindrical container~\cite[][]{CPC}, 
which will be the subject of a future investigation.

Before simulating the patterns, we computed the critical acceleration
$\gamma_c$ from direct numerical simulations (DNS) and compared it
with that of the Floquet linear stability analysis of
\cite{Kumar-jfm-1994}.  These simulations are conducted in a small
periodic domain of length $\lambda_c$ and width $\lambda_c/2$, where
the critical wavelength is $\lambda_c = 2\pi/k_c= 11.3~{\rm mm}$ 
with $k_c$ the critical wavenumber.
We compute the growth rate of the total kinetic energy of the system
(figure \ref{Floquet}(a)) after the growth becomes exponential. Near the
threshold $\gamma_c$, the growth rate varies linearly with $\gamma$,
so that $\gamma_c$ can be deduced by linear interpolation.
Table~\ref{tab_threshold} compares the Floquet predictions of
$\gamma_c$ to the results from DNS for three vibration frequencies.
The discrepancy is under 3\% in all three cases. The ratio of
gravitational to capillary forces is measured by the Bond number,
whose value here is $g\Delta\rho/(\sigma k_c^2)=1.5$, indicating that
both forces are at work.

\begin{figure}
\begin{center}
%\large{a)}\includegraphics*[width=0.45\linewidth]{Subdomains2}\hfill
%\large{b)}\includegraphics*[width=0.45\linewidth]{SubdomainsCyl}
\includegraphics*[width=\columnwidth]{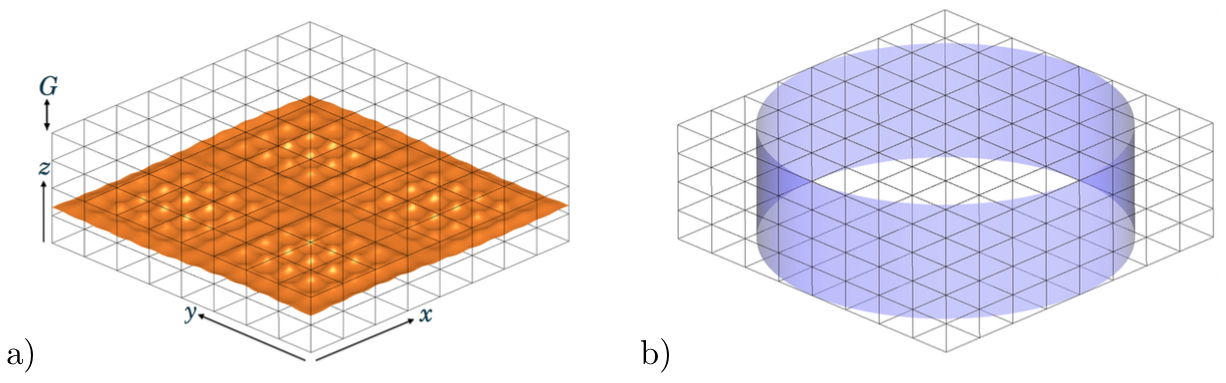}
\end{center}
\caption{\label{domains} Calculation domain of size $L \times L \times L/3$ ($L=132$ mm), divided into $8\times8\times4$ subdomains. a) Interface profile. b) an immersed solid cylinder of diameter $D=L$ and height $h=L/3$ surrounding the flow. The flow outside of the cylinder is not computed. For both (a) and (b) the Cartesian resolution per subdomain is $64^3$, which gives a global resolution of $512\times512\times256$.}
\end{figure}

Our large-scale simulations are carried out with a vibration
frequency of $\omega/2\pi=30$ Hz, i.e. a period of $T\approx0.033$ s,
and an amplitude of $\gamma= g = 1.36\gamma_c$.
Figure \ref{Floquet}(b) tracks the timestep bounds (\ref{eq:bounds})
during the simulation.
It can be seen that $\Delta t_{\rm cap}$ is the limiting timestep
throughout the simulation and is on the order of $5 \times 10^{-5}~{\rm s}$.
The domain is of size $L \times L \times L/3$,
where $L=132~{\rm mm} = 11.7~\lambda_c$.
This domain is subdivided into $8\times8\times4=256$ subdomains whose traces can
be seen in figure \ref{domains}(a), each of which contains $64^3$
gridpoints, leading to an overall resolution of $512 \times 512 \times
256$ gridpoints on a Cartesian mesh. The critical wavelength is thus resolved by $44=512\lambda_c/L$ grid cells. Each subdomain is assigned to a
process thread. 
Our initial condition has zero velocity and random noise on the 
interface of order $10^{-2}$ mm. 
The pattern emerges after approximately 40 forcing periods, 
increases exponentially, and saturates at about 90 forcing periods, 
requiring about 15 days of computation time 
on 256 threads of an IBM x3750-M4.
The cylindrical container has been modeled by an
impermeable solid object as shown in figure \ref{domains}(b).

\section{Results and discussion \label{Resu_General}}

\subsection{Visualisations}

\begin{figure*}
%\large{($i$)}\includegraphics[width=0.29\linewidth,height=0.29\linewidth]{Vz_Dirichlet}\hspace{1.5mm}\large{($ii$)}\includegraphics[width=0.29\linewidth,height=0.29\linewidth]{Vz_Neumann}\large{($iii$)}\includegraphics[width=0.29\linewidth,height=0.29\linewidth]{Vz_Periodic}
\includegraphics[width=\columnwidth]{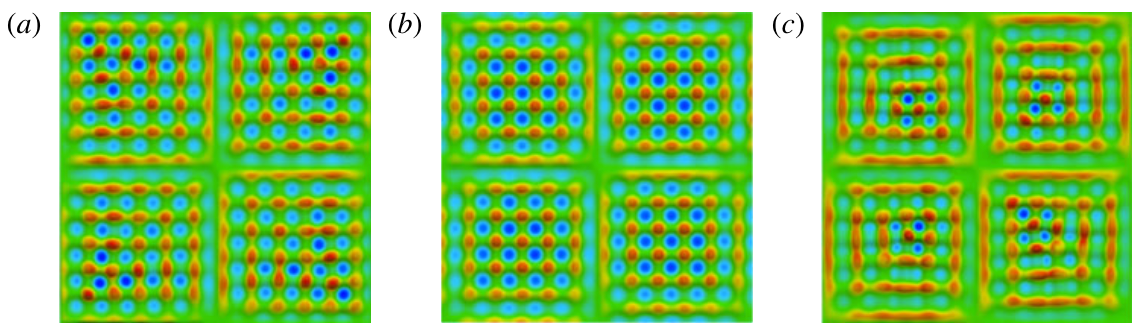}
\caption{\label{BC}
Snapshots of the top view of the interface colored by the vertical velocity for
{\it (a)} no-slip, {\it (b)} free-slip, and {\it (c)} periodic
boundary conditions.}
\end{figure*}
\begin{figure*}
\begin{center}
%\large{a)}~\includegraphics[width=0.3\linewidth,height=0.3\linewidth]{INT_726.png}~\large{b)}~\includegraphics[width=0.3\linewidth,height=0.3\linewidth]{INT_736.png}~\large{c)}~\includegraphics[width=0.3\linewidth,height=0.3\linewidth]{INT_746.png}
\includegraphics[width=\columnwidth]{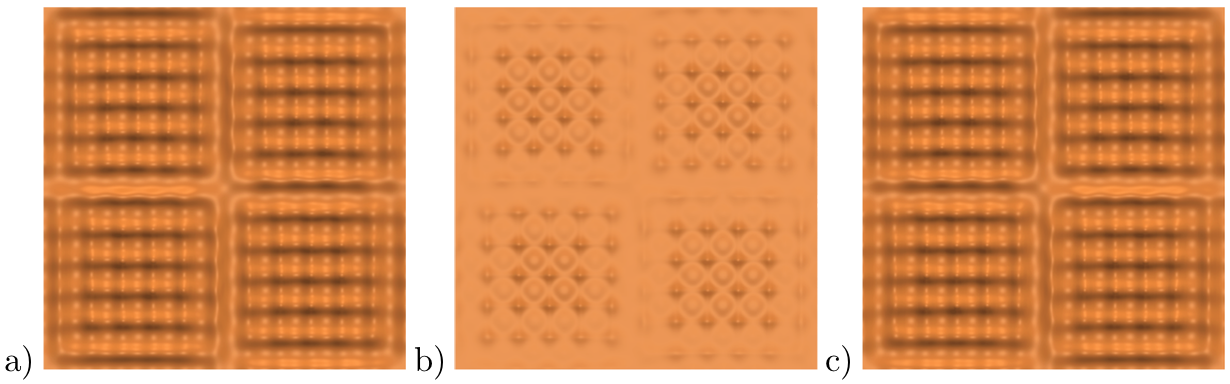}
\end{center}
\caption{\label{Simple_INT_Square}Interface profiles at times separated by intervals of $T/2$, where $T$ is the forcing period and $2T$ is the subharmonic
response period.
In (a) and (c), the small squares consist of wells surrounded by
ridges with peaks at each corner, while in (b), low central peaks are
surrounded by four higher peaks.  In (a), a weak diagonal ``bridge''
connects the large squares on the bottom left and top right; in (c),
the bridge links the top left and bottom right.}
%\end{figure*}
%
%\begin{figure*}
%\begin{center}
%\large{a)}~\includegraphics[width=0.99\linewidth]{Glyph_726}
%\large{b)}~\includegraphics[width=0.99\linewidth]{Glyph_736}
%\large{c)}~\includegraphics[width=0.99\linewidth]{Glyph_746}
%\end{center}
\includegraphics[width=\columnwidth]{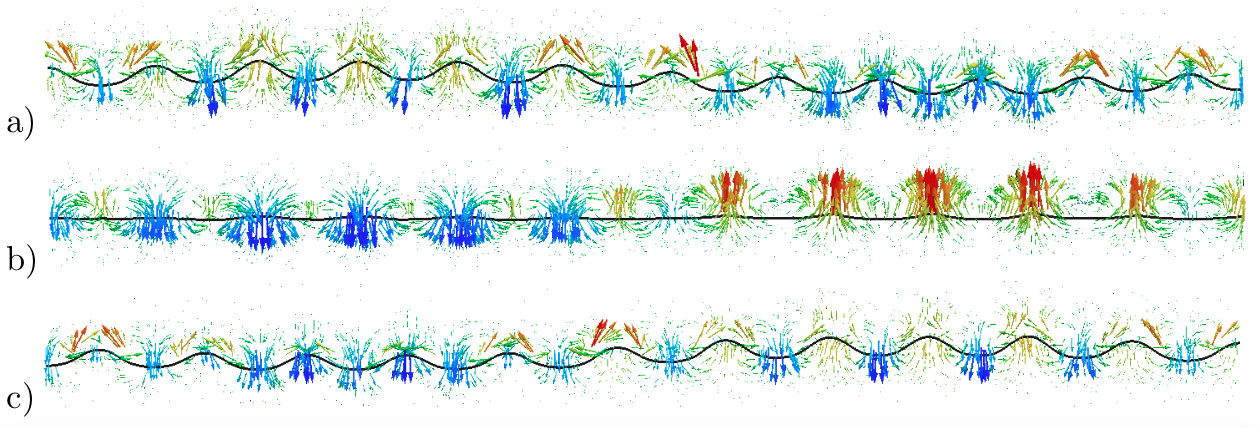}
\caption{\label{velocities}Velocity profiles on the plane $y=3L/4$, corresponding to each of the three instants shown in figure \ref{Simple_INT_Square}.
The vectors are colored according to their vertical component.}
\end{figure*}

The top and bottom of the domain are taken to be rigid; we impose
no-slip boundary conditions on the velocity field. Three types of
boundary conditions have been tested on the lateral walls, {\it (a)}
no-slip, {\it (b)} free-slip, and {\it (c)} periodic.
Although our parameters were originally chosen for a different 
purpose, we find that the patterns that we observe 
are strikingly similar to those observed in earlier experiments by 
\cite{Douady-epl-1988}, figure 1(b) and \cite{Douady-jfm-1990}, 
figures 10(b), 11(b) and 15(b), with different parameters:
water, with $\rho_1 =1000\;{\rm kg}\;{\rm m}^{-3}$, 
$\mu_1=9\times 10^{-4}\;{\rm kg}\;{\rm m}^{-1}\;{\rm s}^{-1}$, 
and $\sigma = 0.03\;{\rm kg}\;{\rm s}^{-2}$, 
in a container of dimensions 
$80\;{\rm mm}\times 80\;{\rm mm}\times 5\;{\rm mm}$
oscillated with $f=37\;{\rm Hz}$ or $f=70\;{\rm Hz}$.

For all three boundary conditions, we observe superlattice-like
patterns shown in figure \ref{BC}. 
The interface in the $x$-$y$ plane consists of four large square sub-blocks, 
each composed of smaller squares whose sides have 
approximate length $\lambda_c$.
In each case, the overall pattern has symmetry $D_2$, i.e., it is 
invariant under reflections through the two diagonals. 
Each of the four sub-blocks is in phase opposition with its two adjacent
neighbors.
%, so that the pattern is invariant under the combined operations 
%of rotation by $\pi/2$ and the inversion of crests and troughs. 
The patterns in (a) and (c) have periodicity length $L$ 
while (b) has a period of $2L$.
Although one would expect the periodic case (c) to be the simplest to 
analyze, we observe that the blocks in the free-slip case (b) 
are more homogeneous than those in cases (a) and (c); 
we will therefore focus on this case.

Figure \ref{Simple_INT_Square} displays the temporal evolution of the
interface at saturation.
(A movie of this evolution is available as supplementary material
to this article.)
After one period of forcing oscillation $T$, troughs are replaced by crests 
(see figure \ref{Simple_INT_Square}(a) and (c)) 
as a consequence of the subharmonic behavior of the interface. 
This subharmonic behavior is also displayed by the large sub-blocks,
since the appearance of each block is transformed into that of its two
adjacent neighbors after time $T$ and then returns to its initial
shape after time $2T$ has passed.  The pattern is thus invariant under
the combined operations of rotation by $\pi/2$ and time-translation by
$T$, a spatio-temporal symmetry.

Figure \ref{velocities} shows the interface and the velocity on a 
vertical slice at the same three instants. 
The contrasting behavior of the velocity in the left and right halves 
of the slice is displayed. 
The interface of figure \ref{velocities}(a) has a maximum (minimum) 
at the left (right) boundary and vice versa for (c), illustrating again 
that the field is not periodic in a domain of length $L$.
%The amplitude of the velocity is greatest in figure \ref{velocities}(b), 
%when the interface is least perturbed. 

\subsection{Fourier spectra}

\begin{figure*}
\includegraphics[width=\columnwidth]{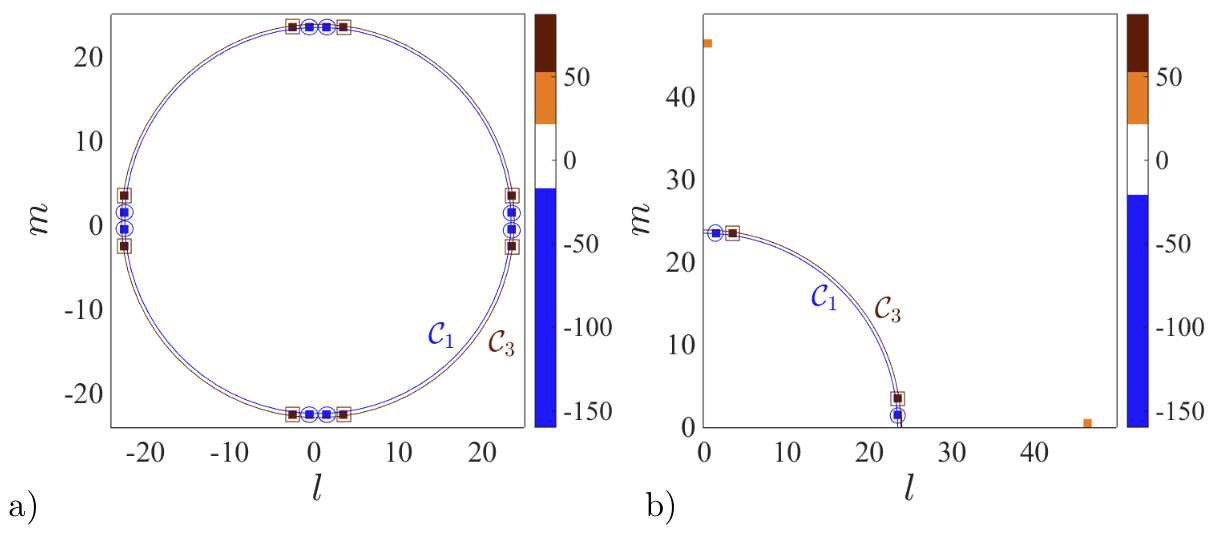}
\caption{Spatial spectrum of a snapshot of
     the interface in the doubled domain.
     Circles $\mathcal{C}_1$ (dark blue) and $\mathcal{C}_3$ (dark brown)
     have slightly different radii. $\mathcal{C}_1$ contains
     the fundamental mode $\hat\zeta_{23,1}$ and its images under
     reflection and rotation (dark blue dots
     surrounded by circles). $\mathcal{C}_3$ contains the highest
     third-harmonic mode $\hat\zeta_{23,3}$ and its images
     (dark brown dots surrounded by squares).
     Only these modes and, to a lesser extent, the weaker second
     harmonic $\hat\zeta_{46,0}$ (orange) have significant amplitude.
     a) Restriction to $[-25,25]^2$ highlights the square symmetry of the
     pattern. b) Restriction to the first
     quadrant $[0,50]^2$ includes $\hat{\zeta}_{46,0}$.}
   \label{fig-spatial_spectrum}
\end{figure*}

For free-slip boundary conditions, the pattern is periodic when it is
doubled horizontally by reflection at the boundaries.  
The doubled pattern is invariant under the symmetry group $D_4$ 
of the square, consisting of reflections through the $x$ and $y$ coordinate 
axes and rotations of multiples of $\pi/2$.
We decompose the doubled interface height profile $\zeta(x,y,t)$ 
into spatial Fourier modes, and then into spatio-temporal modes as follows:
\begin{equation}
\zeta(x,y,t)=\sum_{l,m=-\infty}^{\infty}
\hat{\zeta}_{l,m}(t)e^{i\textbf{k}_{l,m}\cdot\textbf{x}}=\sum_{l,m=-\infty}^{\infty}
\left[\sum_{n=-\infty}^{\infty}\hat{\hat{\zeta}}_{l,m,n}e^{in\Omega t/2}\right]
e^{i\textbf{k}_{l,m}\cdot\textbf{x}}
\label{eq:temporal}
\end{equation}
where $\textbf{k}_{l,m}\equiv\left(\frac{\pi l}{L},\frac{\pi m}{L}\right)$, $\textbf{x}=\left(x,y\right)$ and $k_{l,m}=|\textbf{k}_{l,m}|$.
The square symmetry of the doubled interface implies that the eight modes
in an octet $\hat{\zeta}_{\pm l,\pm m}$ and $\hat{\zeta}_{\pm m,\pm l}$
all have the same amplitude.
Figure \ref{fig-spatial_spectrum} shows two such octets,
with representative elements $\hat\zeta_{23,1}$ and $\hat\zeta_{23,3}$,
located on circles which we will call $\mathcal{C}_1$ and $\mathcal{C}_3$,
respectively, i.e. with $l=23$ and $m=1$ or $m=3$.

The circle $\mathcal{C}_1$ has radius $k_c$, the critical wavenumber. 
The dominant wavevector octet of our pattern,
represented by $\textbf{k}_{23,1}$, is in accordance with 
the experimental observations of \cite{Douady-epl-1988},
who scanned over the forcing frequency in order to
vary the selected wavenumbers and patterns.
They noted that the range of existence was greatest for 
pairs $(l,m)$ with a small ratio $m/l$ and was maximum for $m=1$,
as in our case, i.e. for wavevectors which are almost parallel to the walls. 

The modes on $\mathcal{C}_3$ can be seen as the result of
cubic interactions between modes of $\mathcal{C}_1$:
\begin{equation}\label{eq-3resonance}
\begin{array}{c}
{\bf k}_{23,3}={\bf k}_{23,1}+{\bf k}_{23,1}+{\bf k}_{-23,1}
\end{array}
\end{equation}
Modes on $\mathcal{C}_3$ are fed continually and damped slowly, if at all, 
since $|k_{23,3}-k_c|/k_c = 0.75\%$. 
Generalizing (\ref{eq-3resonance}) to any $m \ll l$
shows that such cubic interactions lead to $\textbf{k}_{l,3m}$,
with $|k_{l,3m}-k_c|/k_c \sim 4(m/l)^2$. 
Due to its proximity to the critical circle, 
this mode is weakly damped and should often be present when 
Faraday waves are excited in a square domain which is
much larger than the critical wavelength. 
%
%The next closest wavenumber is generated by the cubic interaction
%${\bf k}_{l\pm 2m,m}={\bf k}_{l,m}+{\bf k}_{m,l}+{\bf k}_{m,-l}$,
%which satisfies $|k_{l\pm 2m,m}-k_c|/k_c \sim \epsilon$.
%Thus, a superlattice pattern composed of
%$\textbf{k}_{l,m}$ and $\textbf{k}_{l,m+2}$ should
%occur often when Faraday waves are excited in a square domain.
%
A second harmonic of $\mathcal{C}_1$, arising from the quadratic interaction
\begin{equation}
  {\bf k}_{46,0} = {\bf k}_{23,1} + {\bf k}_{23,-1} 
\end{equation}
is seen in figure \ref{fig-spatial_spectrum}(b)
along with its image under $\pi/2$ rotation.
These modes occur in a quartet rather than an octet because of their
reflection symmetry in the coordinate axes 
and are of smaller amplitude than the modes on $\mathcal{C}_1$
or $\mathcal{C}_3$. 
The experimental techniques of \cite{Douady-epl-1988} did
not give them access to amplitudes of Fourier modes other
than the dominant one.

%and had a smaller threshold in amplitude than those in which $l\approx m$.
%and hypothesized that this is due to
%the different dissipation of these modes at rigid boundaries. 
%This is possible for rigid boundaries

According to the terminology of \cite{Simonelli-jfm-1989}, 
the four components $\hat{\zeta}_{\pm l,\pm m}$ comprise one pure mode,
while $\hat{\zeta}_{\pm m,\pm l}$ comprise another pure mode.
\cite{Douady-epl-1988} observed mixed modes (which they called symmetric)
for small $m/l$, and pure modes (which they called dissymmetric)
for larger $m/l$.
Pure modes may be combined to form two types of mixed modes, the
in-phase mode in which all of the components have the same sign (as in
our figure \ref{fig-spatial_spectrum}), and the out-of-phase mode in which
the components of the two pure modes are of opposite sign.  
\cite{Silber-phl-1989} showed that these two types of mixed modes 
are equivalent if $l$ and $m$ have opposite parities. 
\cite{Crawford-physd-1991} showed that this was also true 
for $l$ and $m$ with the same parity (as in our case)
by invoking the concept of hidden symmetries. 
In a square container with Neumann boundary conditions, the doubled 
pattern formed by reflecting at each boundary, as we have done above, 
satisfies periodic boundary conditions, whose 
inherited (hidden) translation invariances
are incorporated into the normal form.

\begin{figure*}
%\includegraphics[height=0.36\linewidth]{pattern123_simple_domain}
%\includegraphics[height=0.36\linewidth]{pattern323_simple_domain}
%\includegraphics[height=0.36\linewidth]{reconstruction_simple_domain}
%\hspace{-\columnwidth}\large{a)}
%\hspace{0.33\columnwidth}\large{b)}\hspace{0.3\columnwidth}\large{c)}
\includegraphics[width=\columnwidth]{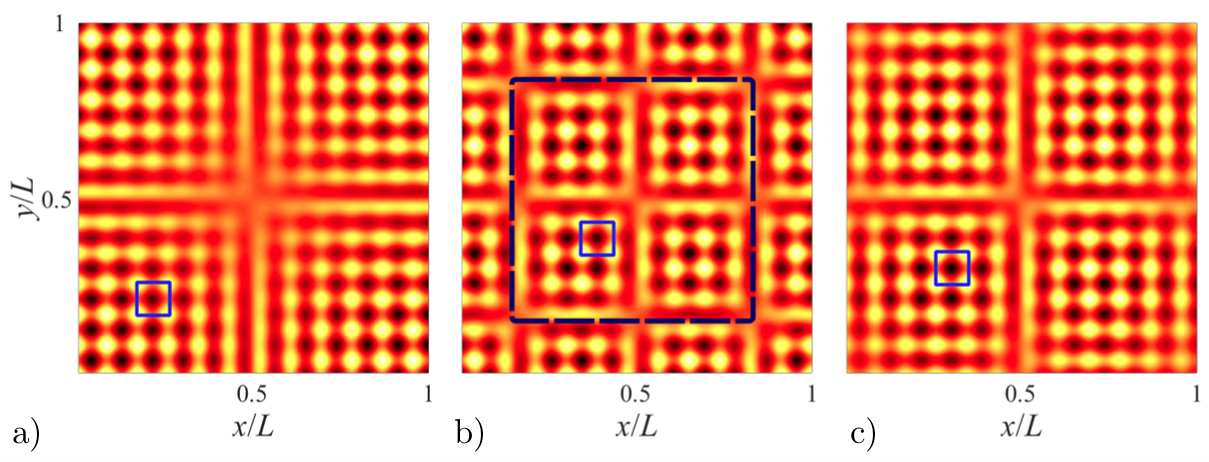}
    \caption{Patterns constructed from Fourier modes
     (a) in $\mathcal{C}_1$, (b) in $\mathcal{C}_3$, and
     (c) a superposition of the two.
     Lighter (darker) zones correspond to the peaks (troughs) of the interface.
     Superlattices, consisting of smaller and larger squares,
     are observed in all of the plots,
     Smallest boxes are approximately of size $\lambda_c$.
     b) Medium squares of size $2L/3$ are present (long-dashed box).
     %Careful examination of consecutive medium squares shows their differences.
     c) Reconstructed pattern, made by combining patterns in
     (a) and (b) weighted by the corresponding
     Fourier coefficients $\hat{\zeta}_{23,1}$ and $\hat{\zeta}_{23,3}$.
     Compare with figure~\ref{Simple_INT_Square}(c). }
   \label{fig-simple_superlattices}
\end{figure*}

\begin{figure*}
%     \large{a)}~\includegraphics[width=0.47\linewidth]{modes_k123_k323_t_plus_k046}\hfill
%     \large{b)}~\includegraphics[width=0.47\linewidth]{TFT_modes_k123_k323_t_plus_k046}
\includegraphics[width=\columnwidth]{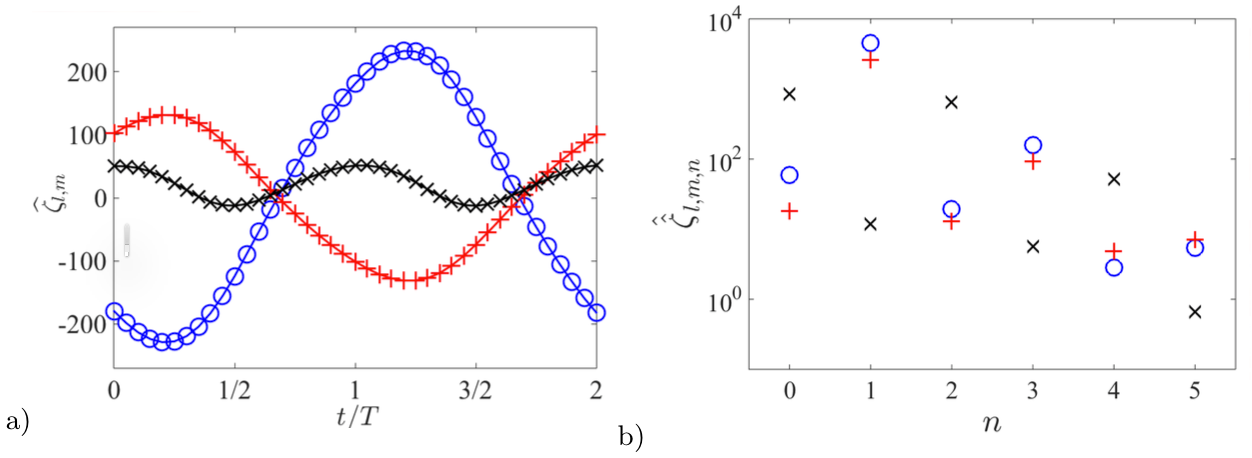}
   \caption{a) Temporal evolution of the three main spatial modes.
    Modes $\hat\zeta_{23,1}$ (blue curve and circles) and $\hat\zeta_{23,3}$
(red curve and $+$ symbols) are subharmonic: their period of oscillation is
twice that of the forcing. Mode $\hat\zeta_{46,0}$ (black curve and $\times$ symbols) is harmonic.
     b) Temporal spectrum of the main modes. $\hat\zeta_{23,1}$,
     $\hat\zeta_{23,3}$ contain only odd temporal
     harmonics while $\hat\zeta_{46,0}$ has mainly even harmonics.
     Each spatial mode has a finite $n=0$ component and hence a non-zero temporal mean.}
   \label{fig-temporal_evolution}
\end{figure*}

For each of $\mathcal{C}_1$ or $\mathcal{C}_3$,
the resulting pattern has a superlattice structure,
in which squares of different sizes coexist,
as shown in figure \ref{fig-simple_superlattices}(a) and (b).
Although shown in domains of length $L$, the periodicity length is $2L$.
The smallest squares have length scale $2\pi/k_{23,m}$ for $m=1,3$.
The patterns formed by modes on $\mathcal{C}_3$ have an additional
intermediate length scale $2L/3$.
Close examination of figure \ref{fig-simple_superlattices}(b)
shows that adjacent medium squares are almost, but not exactly identical.
Figure \ref{fig-simple_superlattices}(c) shows a superposition of
the two patterns of figure \ref{fig-simple_superlattices}(a) and (b),
weighted by the amplitudes of their respective Fourier components.
During most of the time, the interface resembles this superposition;
compare with figure \ref{Simple_INT_Square}.
When the amplitudes of modes on $\mathcal{C}_1$ and $\mathcal{C}_3$
approach zero, which occurs twice during one subharmonic period (see
figure \ref{fig-temporal_evolution}(a) and figure \ref{Simple_INT_Square}(b)),
it is necessary to take modes like $\hat{\zeta}_{46,0}$ into account. 

%\begin{figure*}
%\hspace{-0.05\linewidth}\large{a)}~\includegraphics[height=0.47\linewidth]{reconstruction_pix}\hspace{-0.05\linewidth}
%\large{b)}~\includegraphics[height=0.47\linewidth]{double_pattern_max_bis}
%   \label{fig-reconstruction}
%\end{figure*}

The temporal evolution of $\hat\zeta_{23,1}$, $\hat\zeta_{23,3}$ and
$\hat\zeta_{46,0}$ is displayed in figure \ref{fig-temporal_evolution}(a)
and the spatio-temporal spectrum is shown in
figure \ref{fig-temporal_evolution}(b).
The mode amplitudes $\hat\zeta_{23,1}$ and $\hat\zeta_{23,3}$ are related
by a multiplicative constant and contain only subharmonic temporal components
(i.e. $n$ odd in (\ref{eq:temporal})).
The temporal component $n=1$ corresponding to
$\Omega=\omega/2$ dominates the temporal evolution. Both
$\hat\zeta_{23,1}$ and $\hat\zeta_{23,3}$ evolve in phase opposition, crossing zero at the same time and yielding patterns such as that in figure \ref{Simple_INT_Square}(b). In contrast, $\hat\zeta_{46,0}$
is harmonic and dominated by two temporal components $n=0$ and $n=2$,
as expected from the quadratic
resonance of $\hat\zeta_{23,1}$ with itself or with $\hat\zeta_{23,3}$.
Each spatial mode has a finite $n=0$ mean component;
the offset of $\hat\zeta_{46,0}$ is especially noticeable.

\subsection{Simulation in a cylindrical container}

\begin{figure*}
\begin{center}
%\large{a)}~\includegraphics[width=0.3\linewidth,height=0.3\linewidth]{Cyl_200.png}~\large{b)}~\includegraphics[width=0.3\linewidth,height=0.3\linewidth]{Cyl_210.png}~\large{c)}~\includegraphics[width=0.3\linewidth,height=0.3\linewidth]{Cyl_220.png}
\includegraphics[width=\columnwidth]{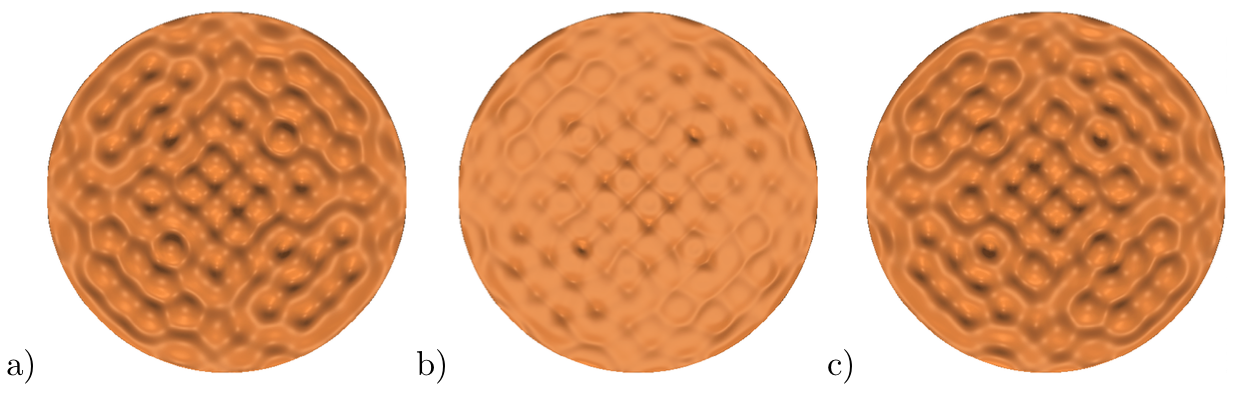}
\end{center}
\caption{\label{Simple_INT_Cylinder}Interface profile in the cylinder at
time intervals of $T/2$. The central part of the domain is still occupied
by squares, but five-sided cells are present near the boundaries,
allowing the pattern to fit inside a circle.}
\end{figure*}

We performed simulations under the same conditions, but in a
cylindrical container of diameter $D=L$ and height $h=L/3$ (see
figure \ref{domains}(b)).  
The lateral boundary conditions are free-slip and the advancing and receding 
contact angles are fixed at $100^{\circ}$ and $60^{\circ}$ respectively. 
Figure \ref{Simple_INT_Cylinder} shows snapshots of the
interface every $T/2$. 
(A movie of this evolution is available as supplementary material
to this article.) 
The pattern has the spatio-temporal symmetry of invariance under 
combined time-evolution by $T/2$ and rotation by $\pi/2$, while 
each instantaneous pattern is reflection-symmetric about both diagonal axes.
At the center, we observe squares whose characteristic wavelength
remains close to $\lambda_c$ of Table \ref{tab_threshold} for $f=30\;{\rm Hz}$.
Closer to the boundary, five-sided cells can be seen as the pattern 
organizes itself to fit in the circular domain. 
Aside from this, the pattern is relatively insensitive to 
the circular shape of the domain because $L/\lambda_c = 11.7 \gg 1$, 
in contrast with small-radius experiments and theory. In most of 
these \cite[e.g.][]{Ciliberto-jfm-1985,Crawford-physd-1990,Das-jfm-2008,Batson-jfm-2013}, 
the interface takes the form of Bessel functions, 
including axisymmetric modes, while \cite{Rajchenbach-prl-2013} 
report intriguing patterns of the form of stars and pentagons. 

\section{Conclusion}

We have presented the first three-dimensional numerical results of the
Faraday instability in domains much larger than the critical
wavelength in both square and cylindrical containers. In the square
domain, the interface displays patterns that are very similar to those
found in the work of \cite{Douady-epl-1988} and
\cite{Douady-jfm-1990}.
By means of a spatial spectral decomposition,
we have described these patterns quantitatively and have found that
they have a complex superlattice-like structure due to resonances
between modes of very similar wavelengths.
We conjecture that this pattern arises from two effects:
(i) In a square container which is large compared to the critical wavelength,
Faraday wave patterns tend to be mixed modes whose wavevectors
are almost parallel to the boundaries of the box, 
as stated by \cite{Douady-epl-1988}. 
(ii) In this case, cubic nonlinearities generate another wavelength 
which is very close to the dominant one.
The combination of the two effects leads to the superlattice
pattern that we observe in our simulations and we infer
that such patterns should be observed in a wide variety of
systems. 
The present work demonstrates that it is now possible to
numerically compute all the fields of Faraday waves in domains 
sufficiently large to produce complex patterns such as superlattices,
quasicrystalline patterns or oscillons. 
The hydrodynamic mechanisms
responsible for the formation of these patterns can then be explored
numerically.

\subsubsection*{Acknowledgements}
We thank S.~Douady, S.~Fauve, and G.~Pucci for helpful discussions. This
work was performed using high performance computing resources provided
by the Institut du Developpement et des Ressources en Informatique
Scientifique (IDRIS) of the Centre National de la Recherche
Scientifique (CNRS), coordinated by GENCI (Grand \'Equipement National
de Calcul Intensif). N.~P\'erinet was supported by a grant from
CONICYT - FONDECYT/postdoctoral research project 3140522.
L.S.T. acknowledges support from the Agence Nationale de la Recherche (ANR)
for the TRANSFLOW project. This
research was supported by the Basic Science Research Program through
the National Research Foundation of Korea (NRF) funded by the Ministry
of Science, ICT and future planning (NRF-2014R1A2A1A11051346).

\bibliographystyle{jfm}

\end{document}